\documentclass[conference]{IEEEtran}
\usepackage{cite}
\usepackage{amsmath,amssymb,amsfonts}
\usepackage{algorithmic}
\usepackage{graphicx}
\usepackage{textcomp}
\usepackage{xcolor}

\usepackage{csquotes}

\usepackage{url}

\usepackage{enumitem}
\newlist{rqs}{enumerate}{1}
\setlist[rqs]{%
    label*=\textbf{RQ\arabic*~},
    resume} %

\def\BibTeX{{\rm B\kern-.05em{\sc i\kern-.025em b}\kern-.08em
    T\kern-.1667em\lower.7ex\hbox{E}\kern-.125emX}}

\usepackage{booktabs}   %
\usepackage{tcolorbox}  %
\usepackage[hidelinks]{hyperref}   %
\usepackage[capitalize]{cleveref}   %

\definecolor{colorA}{RGB}{114,141,196}  %
\definecolor{colorB}{RGB}{214,104,99}   %
\definecolor{colorC}{RGB}{109,179,147}  %
\definecolor{colorD}{RGB}{199,123,227}  %
\definecolor{colorE}{RGB}{255,193,116}  %
\definecolor{colorF}{RGB}{179,140,121}  %

\begin{document}

\title{Evidence Profiles for Validity Threats in Program Comprehension Experiments}

\author{
	\IEEEauthorblockN{Marvin Muñoz Barón, Marvin Wyrich, Daniel Graziotin and Stefan Wagner}
    \IEEEauthorblockA{
    Institute of Software Engineering,
    University of Stuttgart\\
    Stuttgart, Germany\\
    \{firstname.lastname\}@iste.uni-stuttgart.de
    }
}

\maketitle

\begin{abstract}
Searching for clues, gathering evidence, and reviewing case files are all techniques used by criminal investigators to draw sound conclusions and avoid wrongful convictions.
Medicine, too, has a long tradition of evidence-based practice, in which administering a treatment without evidence of its efficacy is considered malpractice.
Similarly, in software engineering (SE) research, we can develop sound methodologies and mitigate threats to validity by basing study design decisions on evidence.

Echoing a recent call for the empirical evaluation of design decisions in program comprehension experiments, we conducted a 2-phases study consisting of systematic literature searches, snowballing, and thematic synthesis.
We found out (1) which validity threat categories are most often discussed in primary studies of code comprehension, and we collected evidence to build (2) the evidence profiles for the three most commonly reported threats to validity.

We discovered that few mentions of validity threats in primary studies (31 of 409) included a reference to supporting evidence.
For the three most commonly mentioned threats, namely the influence of programming experience, program length, and the selected comprehension measures, almost all cited studies (17 of 18) did not meet our criteria for evidence.
We show that for many threats to validity that are currently assumed to be influential across all studies, their actual impact may depend on the design and context of each specific study.

Researchers should discuss threats to validity within the context of their particular study and support their discussions with evidence.
The present paper can be one resource for evidence, and we call for more meta-studies of this type to be conducted, which will then inform design decisions in primary studies.
Further, although we have applied our methodology in the context of program comprehension, our approach can also be used in other SE research areas to enable evidence-based experiment design decisions and meaningful discussions of threats to validity.

\end{abstract}

\begin{IEEEkeywords}
program comprehension, threats to validity, empirical software engineering
\end{IEEEkeywords}

\section{Introduction}

Between 2008 and 2013, drivers on German highways had reasons to be concerned.
Criminal investigators were on the trail of \textit{The Highway Shooter}, as the media called them.
Over the years, the Highway Shooter had fired a gun at passenger cars more than 700 times, with a predilection for trucks\cite{bgh:2015:news}.
As it became clear that the shooter was mobile and that cases had spread across the country, the German Federal Criminal Police Office took over.

Several attempts were made to apprehend the criminal, most of which were covered intensively by the media.
These attempts culminated in a public manhunt with a reward of 100\,000~\texteuro.
In the end, however, what led to success was a combination of two measures: firstly, carriers were sensitized to immediately report bullet holes in their vehicles.
Secondly, several license plate recording devices were set up along the highways.
April 2013 was a hectic month: six shootings were reported within five days.
The culprit's route started showing patterns. 
These patterns, crossed with the captured license plate numbers, ultimately resulted in a single revealing license plate number.
The final evidence led to the arrest of The Highway Shooter in June 2013.

In the same year that the criminal was caught, Wohlin published a paper on evidence profiles for software engineering research and practice~\cite{Wohlin:2013:EvidenceProfile}.
The overlap in time is likely coincidental, but some parallels between the case and Wohlin's proposed methodology are not.

Wohlin~\cite{Wohlin:2013:EvidenceProfile} motivates the work with the increasing importance of evidence-based software engineering (EBSE).
For example, critical decisions such as the introduction of a new tool that could affect software quality or developer productivity should be based on scientific evidence.
Wohlin proposes a model by which evidence from different studies could be evaluated in a manner similar to criminal law investigations.
Practical conclusions could be drawn by putting the individual pieces together in an evidence profile.
Just as with license plates and reports from drivers in the case of The Highway Shooter, we can more effectively provide evidence in software engineering investigations.

Just as investigators were left in the dark for several years in the case of The Highway Shooter, researchers have been uncertain about the consequences of design decisions in code comprehension experiments.
In the corresponding papers, researchers regularly discuss and speculate about the threats to validity of their experiments.
Two meta-studies have categorized these threats, coming up with more than 50 different threat categories~\cite{Siegmund:2015:Confounding,Wyrich:2022:40Years}.
Moreover, papers that do not discuss threats to validity in code comprehension experiments are rather the exception today~\cite{Wyrich:2022:40Years,Schroeter:2017:Comprehending}.
At the same time, hardly anyone is sure about the actual extent of the discussed threats, and almost no paper cites evidence on the assumed threats~\cite{Wyrich:2022:40Years}.
This makes it difficult to evaluate study designs in an evidence-based manner.
Researchers have to decide which of the more than 50 potential threats do, in fact, threaten the validity of a study design, execution, and interpretation, and to which extent they should be disclosed and elaborated on.

In this paper, we apply Wohlin's methodology of evidence profiles.
We echo a recent call by Wyrich et al.~\cite{Wyrich:2022:40Years} to provide more evidence in the design of empirical program comprehension studies.
To that end, we examined the threats to validity in program comprehension experiments to collect evidence of their existence, to understand the context and nature in which they occur, and to ultimately assist researchers in designing controlled experiments with high validity.
Specifically, we extracted the threats to validity reported in 95 code comprehension experiments through thematic synthesis.
Then, focusing on the three most frequently mentioned threat categories, we collected evidence that contradicted or supported the influence of the threat, using systematic literature searches and snowballing.
Finally, we individually scored the evidence that passed our filtering criteria to create an evidence profile, serving as an overview of the evidence for each of the three threat categories.

Our work shows that the extent of commonly assumed threats to validity can be very dependent on the particular design and context factors of a primary study.
Nevertheless, certain threats have so far been generalized to all code comprehension experiments without restriction.
With three evidence profiles, we show how evidence-based experiment design decisions can be made and discussed.
We will illustrate, throughout the paper, how such an approach can improve the validity of experiments even beyond the context of program comprehension for many other experiments in software engineering.

The paper is structured as follows.
In Section~\ref{sec:related_work}, we define the context of our investigation regarding the current handling of threats to validity in software engineering research and explain their specific meaning in the program comprehension field.
In Section~\ref{sec:methodology}, we introduce concrete research questions based on the introductory motivation and describe in detail our methodological approach to answer these questions.
In Section~\ref{sec:results}, we present the results of this work and discuss them in Section~\ref{sec:discussion}, before concluding the paper in Section~\ref{sec:conclusion}.

\section{Background and Related Work}
\label{sec:related_work}

Medicine has a long tradition of evidence-based practice, in which administering a treatment without evidence of its efficacy is considered malpractice.
Sackett et al. describe evidence-based medicine as \enquote{the conscientious, explicit, and judicious use of current best evidence in making decisions about the care of individual patients}~\cite[p. 71]{Sackett:1996:Evidence}.

In software engineering, attempts have been made to establish similar approaches to help practitioners make informed decisions in their daily work~\cite{dyba2005evidence}.
Proponents of evidence-based methods emphasize the importance of using evidence when adopting new technologies and when understanding and identifying problems in existing development processes.
Making uninformed decisions may lead practitioners to favor ineffective solutions over better alternatives, resulting in financial losses or even human costs.
We consider evidence to be the available body of empirical knowledge \enquote{indicating whether a belief or proposition is true or valid}~\cite[p. 607]{Stevenson.2010}.
Throughout our work, this definition and our refinement of the term with different levels of evidence strength in \cref{sec:evidenceprofile} guided us in identifying how empirical knowledge is documented and how we can use it to assist researchers.

Since the first calls for evidence-based software engineering in the early 2000s~\cite{kitchenham2004evidence, dyba2005evidence}, investigative methods from EBSE have found widespread usage.
There has been a significant increase in the number of secondary software engineering studies~\cite{kitchenham2010systematic, budgen2022evolution} and educators are actively incorporating EBSE in their university curricula~\cite{jorgensen2005teaching, rainer2008follow, oates2009using}.
However, recent discussions highlight difficulties in the application of evidence gained from primary and secondary research~\cite{10.1145/2601248.2601274,                                            6681370}.
In response, more tools and structured approaches are being introduced in an attempt to address the slow transfer of knowledge from research to practice (e.g., evidence profiles~\cite{Wohlin:2013:EvidenceProfile}, evidence briefings~\cite{10.1145/2961111.2962603                                            }, and rapid reviews~\cite{cartaxo2020rapid}).

While evidence-based approaches may assist practitioners in making decisions, Kitchenham et al.~\cite{kitchenham2004evidence} also describe how these approaches place additional requirements on researchers when developing experimental protocols.
Ideally, researchers can maximize the range of application of a study's results while minimizing potential threats to validity.

Validity refers to the degree to which we can trust the results of an empirical study~\cite{Kitchenham.2015}.
In our study, a threat to validity refers to deliberate design decisions and uncontrolled extraneous factors that may impair the validity of experimental results.
When readers are explicitly informed about validity threats, they can better assess the context in which experiment results may be applied.
Consequently, they are empowered to understand what difficulties may arise when they attempt to replicate the study design or plan a similar study of their own.

There have been several meta-studies to summarize and categorize common threats to validity in software engineering research.
Petersen and Gencel~\cite{Petersen:2013:Worldviews} compared validity threats with different worldviews.
They assumed that researchers have a subconscious tendency to choose methods based on their worldview.
Likewise, Devanbu et al.~\cite{7886896} found in a case study that programmers tend to hold strong beliefs based on personal experience rather than empirical evidence.
The main implication of these studies is that researchers and practitioners make biased decisions based on their intuition, which is at odds with the main goals of evidence-based methods.
Rather, they should address the discrepancy between the evidence and their perceptions and reconsider their decisions accordingly.
Peterson and Gencel~\cite{Petersen:2013:Worldviews} suggest that the literature needs to be further analyzed regarding the threats and mitigation techniques mentioned therein and that inquiries need to be made into what worldviews dominate in the various sub-disciplines of software engineering.

Biffl et al.~\cite{Biffl.2014} followed this suggestion and created a knowledge base of threats to validity in software engineering experiments to assist researchers in planning their studies.
They found that only a small fraction of validity threats are reported in most studies and that, instead, the vast majority of threats are too specific to be generalized outside the particular research area they occur in.
These findings highlight the complexity of managing threats to validity, as even switching between different sub-disciplines of software engineering introduces a whole new set of potential threats that must be considered.
They conclude that there is a need for an overview of threats to validity as they are reported in specific areas of software engineering research.

Managing threats to validity is particularly difficult in code comprehension experiments, where researchers seek to uncover the underlying processes of how developers understand code and evaluate ways to support that comprehension process scientifically~\cite{601284}.
This complexity is reflected in the wide variety of different methodologies researchers use to address the potential validity threats in their experiments~\cite{Wyrich:2022:40Years}.

For example, Siegmund and Schumann~\cite{Siegmund:2015:Confounding} surveyed the literature to obtain information on confounding parameters in studies of program comprehension.
The main insights they gained were that each paper reported only a small subset of all confounding factors, and that researchers used different methods to mitigate the same factors.
They recommend that other researchers include the identification of confounding parameters in their experimental design and explicitly report the relevant parameters and how their influence is controlled.

Building on the findings of Siegmund and Schumann, Wyrich et al.~\cite{Wyrich:2022:40Years} analyzed the characteristics of 95 source code comprehension experiments and identified several shortcomings and inconsistencies in their experimental designs and in how they reported the threats to validity.
They noted that, currently, researchers tend to rely on intuition when designing their experiments because of how difficult it is to reliably measure a person's understanding of code.
Researchers have to deal with potential threats to validity from over 50 categories and sometimes question whether their measure of understanding is adequate at all~\cite{Wyrich:2022:40Years}.

These meta-studies illustrate the complexity of designing valid code comprehension experiments, considering potential confounding factors and other threats to validity~\cite{Wyrich:2022:40Years,Siegmund:2015:Confounding}.
Rather than choosing methods based on intuition, we argue that researchers should make informed decisions based on empirical evidence.
Previous research has outlined a need for common knowledge bases of validity threats and collections of evidence backing up both the existence of threats and the effectiveness of their mitigation techniques.
The first step in solving these issues is to identify where evidence is lacking.

Surveying the evidence landscape on any particular issue can be quite daunting.
For this purpose, Wohlin~\cite{Wohlin:2013:EvidenceProfile} proposes the creation of evidence profiles to gain an overview of both the amount and the direction of evidence.
By scoring each individual piece of evidence in the context of an explicit research question, researchers are assisted in identifying sufficient or missing evidence.
This approach shares similarities with other meta-studies, such as meta-analyses, in that it aims to summarize and combine the results of multiple studies.
Evidence profiles differ in their method of synthesis, as meta-analyses take a quantitative approach in composing effect sizes, whereas evidence profiles use multiple reviewers to qualitatively evaluate the studies by scoring them.

\section{Method}
\label{sec:methodology}
\begin{figure*}[t]
    \centering
    \includegraphics[width=0.96\textwidth]{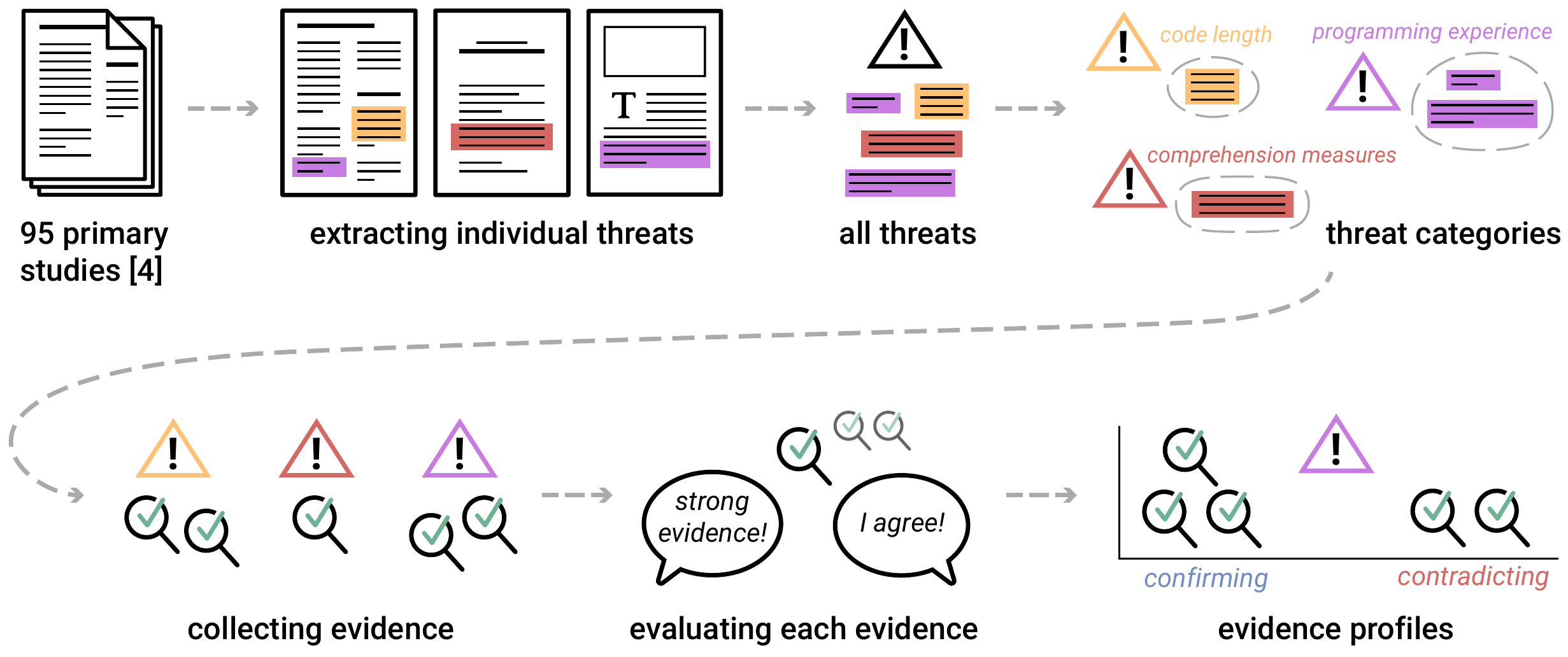}
    \caption{Schematic representation of the research methodology.}
    \label{fig:methodology}
\end{figure*}

We first investigate which threats to validity are most frequently discussed in program comprehension literature (\ref{sec:methods-a}).
The frequency of discussion, however, should only be considered as an indicator of what researchers are most often concerned about, not of the threats' evidence.
Therefore, in a second step, we will examine the scientific evidence for the most frequently discussed threats (\ref{sec:methods-b} and \ref{sec:evidenceprofile}).
\\\cref{fig:methodology} provides a schematic overview of the research methodology.
The research questions that guided us in our endeavor are as follows:

\begin{rqs}[leftmargin=*]
    \item Which validity threat categories are most often discussed in primary studies of code comprehension?
    \item What are the evidence profiles for the three most commonly reported threats to validity in code comprehension experiments?
\end{rqs}

\subsection{Scoping the Relevant Threats}
\label{sec:methods-a}
To answer \textbf{RQ1}, we surveyed existing studies of code comprehension and identified which threats were most frequently mentioned by researchers.
In particular, we extracted the threats to validity reported in 95 source code comprehension experiments found in the openly available dataset by Wyrich et al.~\cite{Wyrich40:dataset}.
Their systematic search protocol included empirical studies of bottom-up code comprehension with human participants, published in a peer-reviewed journal, conference, or workshop before 2020.
While Wyrich et al.~\cite{Wyrich:2022:40Years} reviewed the threats to validity to some extent, we opted to repeat the coding and categorization activities to enable a more fine-grained classification of threats with the explicit goal of gaining insights into the reporting of validity threats as a whole.
Moreover, we performed this bottom-up coding activity to stay closer to the data, building a foundation for the subsequent evidence searches.

We examined the list of primary papers, categorizing and summarizing each threat to validity reported in the full text.
Specifically, we adopted a thematic synthesis~\cite{Cruzes.2011} approach to identify the individual threats to validity and, if given, the mitigation techniques mentioned by the study.
For this, we extracted relevant text areas using inductive coding~\cite{Corbin.2008} to find and describe the passage.
In inductive coding, codes are formulated during the review process as the corresponding concepts become evident.
Throughout the coding process, these codes are refined and reapplied, improving their quality through iteration.
Upon the completion of code assignment, we categorized threat codes into high-level categories and themes.
Further, in some cases, papers already contained evidence of a threat in the form of references to other studies and consequently, those were also documented.
Finally, we counted the number of papers that cite a threat for each threat category, resulting in an overview of the most frequently discussed threat categories.
\newpage
In summary, reviewing threats reported in primary studies consisted of the following individual steps:
\begin{enumerate}
    \item Extraction of relevant text passages on threats to validity, mitigation techniques, and evidence.
    \item Inductive coding, in which each threat is assigned a descriptive code.
    \item Categorization of threat codes and composition into higher-order themes.
    \item Counting the number of papers mentioning each threat category.
\end{enumerate}

\subsection{Evidence Collection}
\label{sec:methods-b}

At this point, we had a list of threat categories, how frequently they were reported, and in some cases, a pool of starting evidence.
Due to the large number of different threat categories, we could not collect evidence for each individual one.
Instead, we focused on the three categories that were mentioned most frequently.
These threat categories were the level of programming experience of participants, the length of the programs used in the experiment, and the measures that were used as a proxy for the concept of comprehension.

To answer \textbf{RQ2}, we conducted systematic searches for each of the three threat categories, collecting evidence on their influence.
The steps described in this section and in \cref{sec:evidenceprofile} were repeated for each individual category.

\subsubsection*{Search Protocol}
Our search protocol utilized four different sources to search for potentially relevant papers.
We describe each of these sources and list the filtering criteria applied to the literature found within them.
Furthermore, we describe how we used snowballing~\cite{Wohlin.2014} as a technique to further extend the search.
Backward snowballing in this context means including the reference list of a paper, while forward snowballing means including papers that cite the paper in question.

\paragraph{Primary Papers}
The 95 papers with primary research on program comprehension may already examine the threat in question as part of their study.
Consequently, we evaluated them as potential evidence.
Forward snowballing was less likely to yield relevant results, as investigating the threat in question was not the main focus of these papers.
Unless the threat was the main subject of a primary study, other studies examining the threat are unlikely to cite the primary study in the context of discussing the threat.
Backward snowballing for this set of papers was not required because the evidence cited in the primary papers was already a separate source, as we describe in the next paragraph.

\paragraph{Evidence Cited in the Primary Papers}
As part of the analysis described in \cref{sec:methods-a}, we identified and documented all references in the primary papers that were cited to support assertions made about a threat to validity.
We analyzed and filtered this list of references in the same manner as the primary papers.
The evidence from this source did not necessarily focus on the threat as their primary object of research.
Forward snowballing was therefore not used, as it was less likely to yield relevant results.
However, we performed backward snowballing, as the papers may refer to similar evidence when comparing their results with other works.

\begin{table*}[t]
    \centering
    \caption[Evidence Types in the Evidence Profile.]{Types of evidence according to the evidence profile.\\ Higher values mean stronger evidence.
    \textcolor{colorA}{Positive values} support the threat, \textcolor{colorB}{negative values} contradict it.}
    \begin{tabular}{p{.04\textwidth} p{.16\textwidth} p{.72\textwidth}}
        \toprule
        Score & Level of Strength & Description \\
        \midrule
        \textcolor{colorA}{+5}/\textcolor{colorB}{-5} & Strong evidence & Studies that focused on the threat in question as the main subject of their investigation or conducted an in-depth analysis of the threat as part of their overall approach and show significant results. Systematic reviews that examined the threat in question and provided a conclusion. \\
        \textcolor{colorA}{+4}/\textcolor{colorB}{-4} & Evidence & Studies that did not have the threat as their main focus but still included it in their analysis. These studies may have more uncontrolled confounding factors, which is why they should be considered separate from strong evidence. \\
        \textcolor{colorA}{+3}/\textcolor{colorB}{-3} & Circumstantial evidence & Similar to evidence, studies that are considered circumstantial evidence did not have the threat as the main focus of their study. Furthermore, they showed additional methodological shortcomings that reduced the reliability of the study and decreased its strength as evidence in our evaluation. \\
        \textcolor{colorA}{+2}/\textcolor{colorB}{-2} & Third-party claim & Studies that made claims about the threat in question but only provided a slim level of empirical backing for said claim. They may have deferred to other sources of information or given general impressions about the influence of the threat in their study, but provided no dedicated statistical analysis of their own to support their claims. \\
        \textcolor{colorA}{+1}/\textcolor{colorB}{-1} & First- or second-party claim & Studies that made a claim about the threat in question but did not provide any empirical backing. This could have been, for example, references to “common knowledge” or speculation. \\
        \bottomrule
    \end{tabular}
    \label{tab:evidence-types}
\end{table*}

\paragraph{Evidence Found Through the Title Search}
To further enrich the dataset with research from sources independent of the primary papers, we also conducted additional systematic searches.
While the literature search by Wyrich et al.~\cite{Wyrich:2022:40Years} already captured code comprehension experiments up to 2020, we were able to extend the range of our search to additionally include literature published between 2020 and 2022.
Furthermore, we focused this search on studies that mention the particular threat to validity in their title using the following search strings in Google Scholar:\newpage
\begin{itemize}
    \item \textbf{Programming Experience} - allintitle: (experience OR novice OR expert) (code OR software OR program) (understandability OR comprehension OR comprehensibility OR readability OR analyzability OR \enquote{cognitive load})
    \item \textbf{Program Length} - allintitle: (size OR length OR short OR long OR LOC OR "lines of code") (code OR software OR program) (understandability OR comprehension OR comprehensibility OR readability OR analyzability OR \enquote{cognitive load})
    \item \textbf{Comprehension Measures} - allintitle: (measure OR measures OR measurement) (code OR software OR program) (understandability OR comprehension OR comprehensibility OR readability OR analyzability OR \enquote{cognitive load})
\end{itemize}
Each search string was composed by combining terms describing a threat (e.g. experience, novice, expert) with terms related to program comprehension experiments (e.g. code, understandability, comprehension).
We used both backward and forward snowballing, as the studies found in the title search were likely to have the threat in question as the main subject of their research.

\paragraph{Evidence Found Through Snowballing}
By exploiting clusters of related research, snowballing can identify relevant literature with high precision, but tends to miss some papers~\cite{Badampudi.2015}.
Therefore, we use a hybrid approach to complement disadvantages of snowballing with a database search and vice versa.
Backward snowballing was applied to the evidence cited in the primary papers, while backward and forward snowballing were used on the evidence from the title search.
In all three cases, we did one iteration of snowballing.

All papers found in the sources a) to d) were filtered according to the following four inclusion criteria and were only included when all of them were fulfilled:

\begin{itemize}
    \item The paper reports a primary study measuring program comprehension.
    \item The paper reports an analysis of the threat in question.
    \item The paper is published in a peer-reviewed journal or conference proceeding.
    \item The paper's full text is available in English.
\end{itemize}

\subsection{Evidence Profiles}
\label{sec:evidenceprofile}

After collecting all available evidence of a particular threat, we evaluated the evidence itself.
To this end, we employed the evidence profile proposed by Wohlin~\cite{Wohlin:2013:EvidenceProfile}.
This profile is a model for evaluating evidence based on criminal law.
Each piece of evidence is judged individually and rated with a corresponding level of strength, from 1 (lowest) to 5 (highest).
This way, evidence strength is represented by an ordinal scale with no zero.
In addition, the profile distinguishes between positive and negative evidence, with positive evidence supporting the theory in question and negative evidence contradicting it.
It is important to stress that evidence of low quality is not synonymous with negative evidence.
Negative evidence can be of high quality, but it opposes the notion that a threat has an influence on the validity of a primary study.
For example, negative evidence with a score of -5 has the same evidence strength as positive evidence with a score of 5.
Thus, a score close to zero indicates the strength of a piece of evidence is low, while a negative or positive score indicates the outcome of the study.

Table~\ref{tab:evidence-types} describes the different levels of strength for evidence.
Due to the flexible nature of the evidence profile, the descriptions do not match the ones provided by Wohlin word-for-word.
Wohlin emphasizes that the evidence profile should be adapted to the context in which it is used.
For example, quality aspects such as \textit{vested interest} and the \textit{aging of evidence} do not play a major role when collecting evidence on threats to validity, unless the threat pertains to a specific technology or approach that may influence experiment results.
By contrast, the methodological rigor captured in the \textit{quality of evidence} as well as the \textit{relevance of the evidence} is of utmost importance and strongly influenced the descriptions of the different levels of evidence strength.

The placement of a study on a particular level is based on its adherence to the factors described in the level description, as well as on the previously mentioned quality aspects.
Therefore, a study may be placed in a lower category if it has significant shortcomings regarding any of the quality aspects.
The concrete definitions of what each level of strength means were established before the evaluation process began.
Because the evaluation of evidence is a largely subjective process, we involved two researchers in this step.
Once both researchers had scored a piece of evidence, they compared their scores and discussed possible disagreements.
In these discussions, each researcher explained their reasoning for their score, and, together, they decided on a single final score.
Where the initial scores were identical, the final score was set to the same value without discussion.
We documented the score of each study and the motivation behind its placement, which can be used to paint an overall picture of the evidence landscape for each threat.
We documented scoring and agreement in our replication package (Section~\ref{sec:data_availability}).
Based on each evidence profile, we provide conclusive recommendations to researchers.

\section{Results}
\label{sec:results}
We first present the results of extracting and categorizing validity threats from 95 primary studies.
This provides us with an answer to \textbf{RQ1}.
We then answer \textbf{RQ2} by presenting the evidence profiles for the three most frequently reported threats to validity in code comprehension experiments.

\vspace{3pt}
\addcontentsline{toc}{subsection}{Validity Threats in Code Comprehension Experiments}
\begin{enumerate}[leftmargin=*, label=\textbf{RQ1}]
    \item Which validity threat categories are most often discussed in primary studies of code comprehension?
\end{enumerate}

\noindent Among the 95 papers, 81~(85\%) mentioned at least one threat to validity, with 45~(47\%) reporting them in a dedicated section and 33~(35\%) differentiating between different types of validity, such as internal or external validity.
The largest concentration of studies was published in the period from 2012 to 2019, with a total of 63 of the 95 studies~(66\%), while only covering 7 years~(18\%) of the entire 40-year span.
The first study in our set of papers to include a dedicated \enquote{threats to validity} section was a study published in 2005\cite{DuBois.2005}.

In total, we found 409 individual threat mentions.
Out of the 409 threat mentions, 198~(48\%) included an explanation of a possible mitigation technique, but only 31~(8\%) were reported with supporting evidence.
The 31 references to supporting evidence were found in 20 out of 95 studies~(21\%).
The 409 threat mentions were then assigned 215 unique threat codes, which captured the different nuances of how a threat was mentioned in a study.
Multiple threat mentions could receive the same code, which is why the number of unique codes is lower than the number of total threat mentions.
To better analyze related threat codes, we additionally categorized them, which resulted in 81 unique threat categories.
For example, the threat category \textit{Programming Experience} included threat codes such as \enquote{Missing diversity in participants' programming experience leads to limited generalizability} and the opposite code \enquote{Diversity in participants' programming experience confounds treatment effects.}
The two individual threat codes differed in nuance, but they both emphasized the importance of programming experience as a threat to validity.
When we counted the number of threat mentions to prioritize the three most common threat categories, each occurrence of either threat code would increase the category's count.

\Cref{tab:most-common-themes} highlights the threat categories with more than five reported threat mentions and shows the themes to which the categories were assigned.
Both categories and themes are presented with the total number of threat mentions.
All mentions from categories with less than five mentions are counted in \enquote{Other Categories} under their respective themes.
\enquote{Theme: Other} contains mentions for threat codes that did not fit into any given theme.
The three most common threat categories (i.e., \textit{Programming Experience}, \textit{Program Length}, and \textit{Comprehension Measures}) are highlighted in blue.
Overall, most threats were related to the characteristics of the code snippets, the individual factors of the participants, or general threats in experimentation.

\begin{table}[t]
    \centering
    \caption{Number of threat mentions per theme and category for categories with more than five threat mentions.}
    \begin{tabular}{p{0.5\linewidth} r}
        \toprule
        Theme and Category & Count \\
        \midrule
        \textbf{Theme: Code Snippets} & \textbf{112}  \\
        \hspace{1em} \textcolor{colorA}{Program Length} & \textcolor{colorA}{26} \\
        \hspace{1em} Complexity & 16 \\
        \hspace{1em} Code Selection & 13 \\
        \hspace{1em} Programming Language & 9 \\
        \hspace{1em} Synthetic Samples & 9 \\
        \hspace{1em} Familiarity & 6 \\
        \hspace{1em} Other Categories & 33 \\
        \textbf{Theme: Participant Factors} & \textbf{101}  \\
        \hspace{1em} \textcolor{colorA}{Programming Experience} & \textcolor{colorA}{44} \\
        \hspace{1em} Number of Participants & 16 \\
        \hspace{1em} Programming Skills & 10 \\
        \hspace{1em} Other Categories & 31 \\
        \textbf{Theme: Experimentation} & \textbf{89}  \\
        \hspace{1em} Learning Effect & 18 \\
        \hspace{1em} Lab Experiment & 11 \\
        \hspace{1em} Fatigue & 9 \\
        \hspace{1em} Code Presentation & 7 \\
        \hspace{1em} Cheating & 6 \\
        \hspace{1em} Other Categories & 38 \\
        \textbf{Theme: Measurement} & \textbf{67}  \\
        \hspace{1em} \textcolor{colorA}{Comprehension Measures} & \textcolor{colorA}{22} \\
        \hspace{1em} Eye-Tracking & 20 \\
        \hspace{1em} Instrumentation & 9 \\
        \hspace{1em} Other Categories & 16 \\
        \textbf{Theme: Comprehension Tasks} & \textbf{21}  \\
        \hspace{1em} Type of Comprehension Task & 7 \\	
        \hspace{1em} Task Difficulty & 6 \\	
        \hspace{1em} Other Categories & 8 \\
        \textbf{Theme: Data Analysis} & \textbf{14}  \\
        \hspace{1em} Statistics & 10 \\
        \hspace{1em} Other Categories & 4 \\
        \textbf{Theme: Other} & \textbf{5}  \\
        \bottomrule
        \textbf{Total} & \textbf{409}  \\
    \end{tabular}
    \label{tab:most-common-themes}
    \hspace{1em}
\end{table}

\vspace{8pt}
    \begin{tcolorbox}[colback=white,colframe=black, title={\textbf{\textcolor{colorE}{RQ1}: \textsf{Main Findings}}}]
        \begin{itemize}[leftmargin=*]
            \item 85\% of code comprehension experiments report at least one threat to validity.
            \item Only 8\% of threat mentions are supported with referenced evidence.
            \item The three most commonly reported threat categories are \textbf{programming experience}, \textbf{program length}, and \textbf{comprehension measures}.
        \end{itemize}
    \end{tcolorbox}

\newpage
\addcontentsline{toc}{subsection}{Evidence Profiles}
\begin{enumerate}[leftmargin=*, label=\textbf{RQ2}]
    \item What are the evidence profiles for the three most commonly reported threats to validity in code comprehension experiments?
\end{enumerate}

\noindent We selected the three threat categories that were most frequently reported and collected evidence on their influence on different facets of validity.
We provide the complete evidence lists, including rationales for the placement of studies in the different evidence categories of an evidence profile, in our replication package (Section~\ref{sec:data_availability}).

\begin{table}[t]
    \centering
    \caption[Number of papers analyzed in the first evidence collection.]{Overview of the evidence analyzed and filtered in each step of the evidence collection for programming experience.}
    \label{tab:evidence-search-results-1}
    \begin{tabular}{l r r r}
        \toprule
        \textbf{Source} & \textbf{Analyzed} & \textbf{Excluded} & \textbf{Final} \\
        \midrule
        a) Primary Papers & 95 & -52 & 43 \\
        b) Cited in Primary Papers & 13 & -12 & 1 \\
        c) Title Search & 54 & -52 & 2 \\
        d) Snowballing & 276 & -262 & 14 \\
        \midrule
        Evidence Profile & 60 & -11 & 49 \\
        \textbf{Total} & \textbf{438} & \textbf{-389} & \textbf{49} \\
        \bottomrule
    \end{tabular}
\end{table}

\begin{figure}[t]
    \centering
    \includegraphics[width=\linewidth]{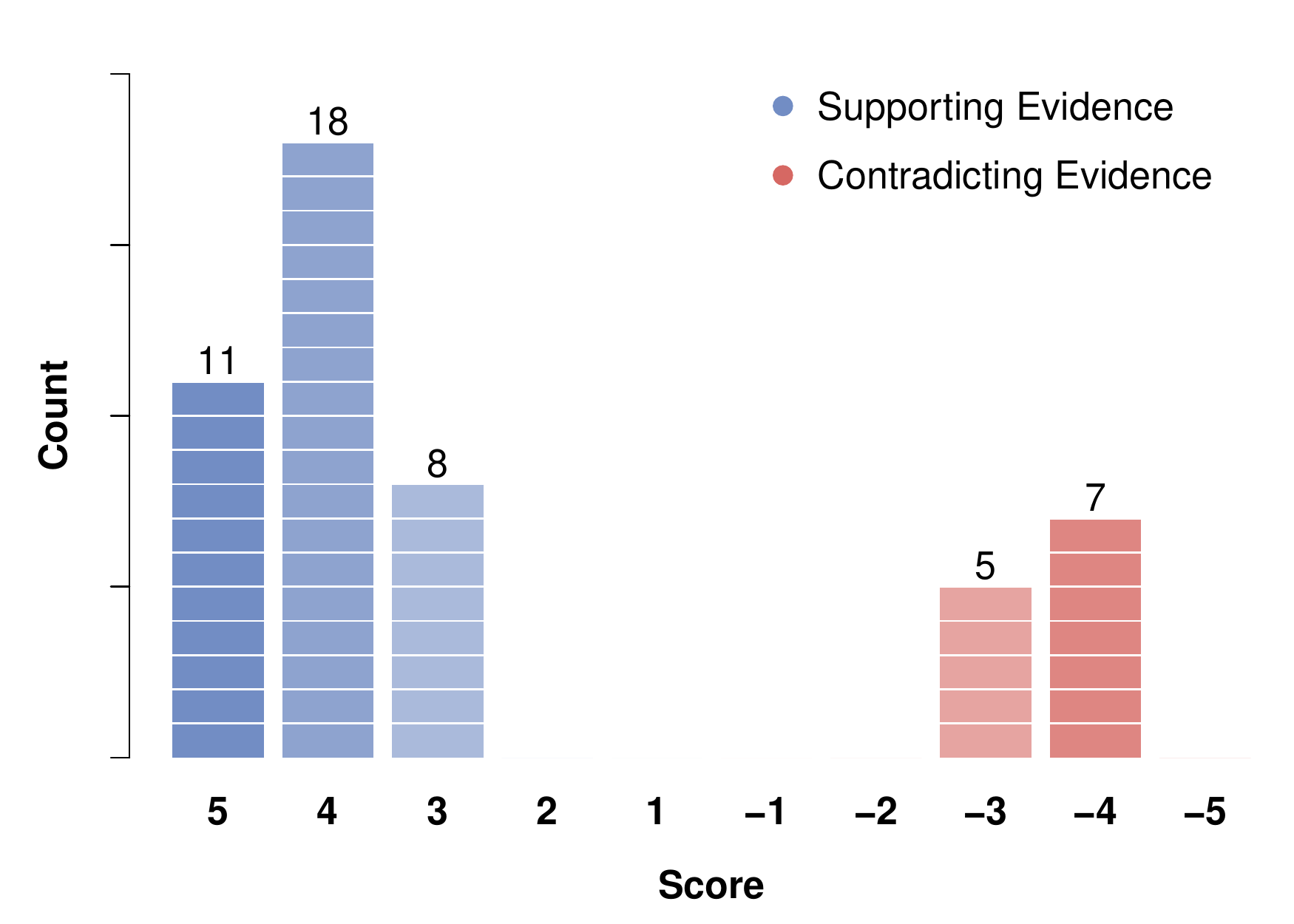}
    \caption{Evidence Profile 1: Programming experience influences code comprehension.}
    \label{fig:evidence-profile-1}
\end{figure}

\subsubsection{Programming Experience}
\label{sec:results-1-experience}
In this analysis, we examine the effect a participant's programming experience has on their code understanding in comprehension experiments.
We collected evidence in the form of studies that measured the programming experience of participants and determined whether it had a significant influence on program comprehension.
\Cref{tab:evidence-search-results-1} shows how many papers we found in each step of the evidence collection and how many we excluded because they did not meet the inclusion criteria.
Overall, most evidence was found in the primary papers, followed by snowballing.
We excluded 12 out of 13 (92\%) documents cited in the references of the primary papers, as they did not meet the filtering criteria.
The reasons for exclusion varied and are described in more detail in Section~\ref{sec:discussion}.

After filtering, 60 papers remained and were evaluated as potential pieces of evidence to be used in the evidence profile.
In this evaluation, 11 additional papers were discarded as they did not meet the criteria to be considered evidence.
For example, one paper was excluded because it was not peer-reviewed and multiple papers were excluded because they either did not measure programming experience or they did not use the collected experience data in their analysis.
\cref{fig:evidence-profile-1} presents the final evidence profile.
The result largely indicates that programming experience influences code comprehension.
In total, 37 (76\%) pieces of evidence were rated as positive evidence and 12 (24\%) were rated as negative evidence.
Furthermore, there were 11 pieces of strong positive evidence and no strong negative evidence.

\subsubsection{Program Length}
\label{sec:results-2-length}
In this analysis, we examine how the length of a program affects program comprehension.
We gather evidence in the form of studies that measure the length of a program and examine its impact on program comprehension.
\Cref{tab:evidence-search-results-2} shows how many papers we found in each step of the evidence collection and how many we excluded because they did not meet the inclusion criteria.
The only evidence was found in the list of primary papers.
All documents cited in the references of the primary papers were excluded as they did not meet the filtering criteria.
Moreover, as we found no relevant papers in b) and c), no snowballing was performed.

\begin{table}[t]
    \centering
    \caption[Number of papers analyzed in the second evidence collection.]{Overview of the evidence analyzed and filtered in each step of the evidence collection for program length.}
    \label{tab:evidence-search-results-2}
    \begin{tabular}{l r r r}
        \toprule
        \textbf{Source} & \textbf{Analyzed} & \textbf{Excluded} & \textbf{Final} \\
        \midrule
        a) Primary Papers & 95 & -78 & 17 \\
        b) Cited in Primary Papers & 3 & -3 & 0 \\
        c) Title Search & 29 & -29 & 0 \\
        d) Snowballing & 0 & 0 & 0 \\
        \midrule
        Evidence Profile & 17 & -4 & 13 \\
        \textbf{Total} & \textbf{127} & \textbf{-114} & \textbf{13} \\
        \bottomrule
    \end{tabular}
\end{table}

\begin{figure}[t]
    \centering
    \includegraphics[width=\linewidth]{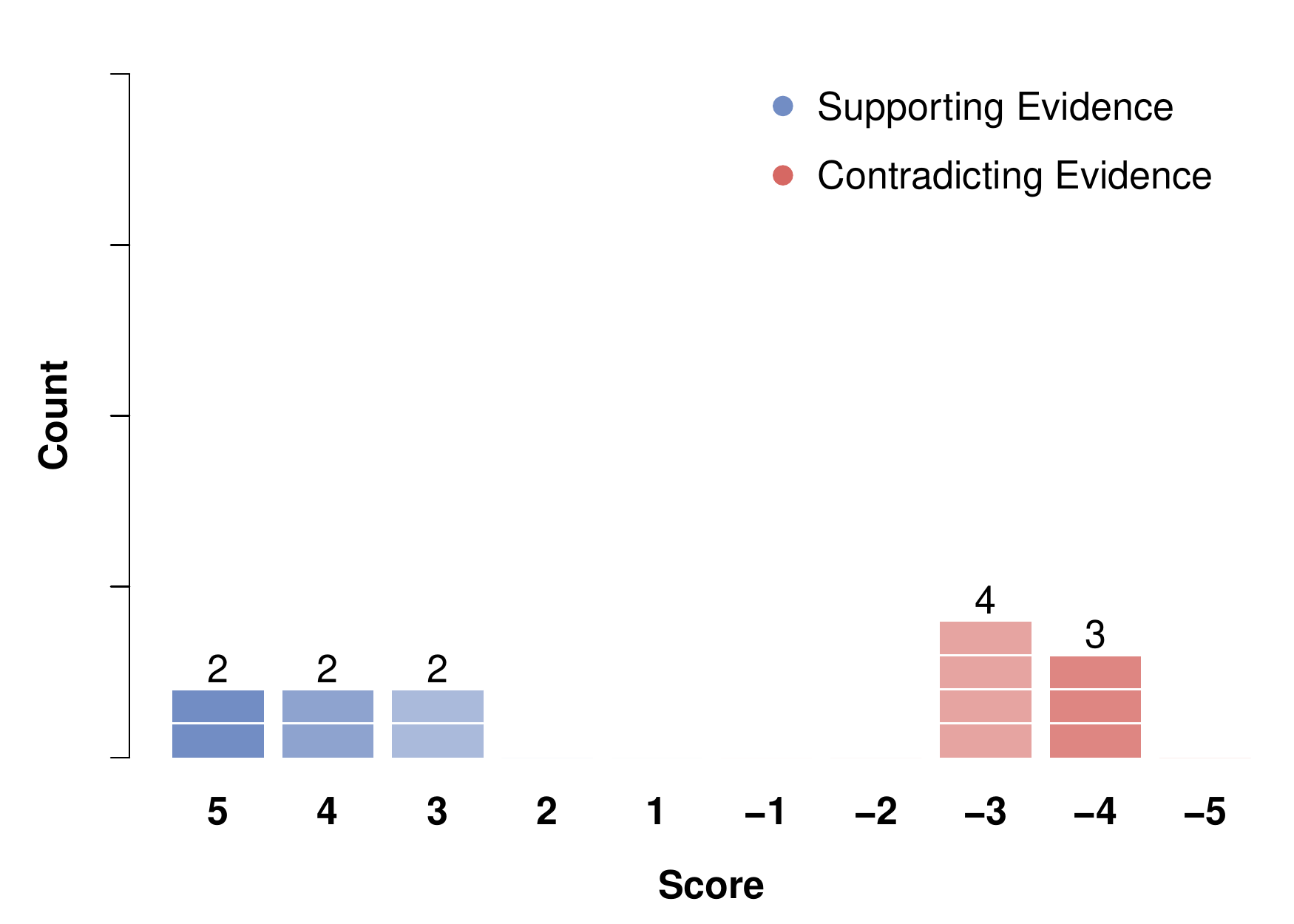}
    \caption{Evidence Profile 2: Program length influences code comprehension.}
    \label{fig:evidence-profile-2}
\end{figure}

After filtering, 17 papers remained and were evaluated as potential evidence using the evidence profile.
In this evaluation, 4 more papers were discarded as they did not meet the criteria to be considered evidence.
The final evidence profile is shown in \cref{fig:evidence-profile-2}.
We found conflicting results regarding the influence of program length on program comprehension: 6 (46\%) pieces of evidence were rated as positive evidence and 7 (54\%) were rated as negative evidence.
There were 2 pieces of strong positive evidence and no strong negative evidence.

\subsubsection{Comprehension Measures}
\label{sec:results-3-measures}
Finally, we examine whether common comprehension measures are associated with distinct aspects of comprehension.
We gather evidence in the form of comparative studies that analyze correlations between commonly used comprehension measures.
Positive evidence describes that different comprehension measures are not correlated and if different measures were used in a primary study, one may arrive at a different conclusion.
\Cref{tab:evidence-search-results-3} shows how many papers were found in each step of the evidence collection and how many were excluded because they did not meet the inclusion criteria.
Overall, we found most evidence in the primary papers, with slightly less evidence found in both the title search and through snowballing.
All documents cited in the references of the primary papers were excluded as they did not meet the filtering criteria.

\begin{table}[t]
    \centering
    \caption[Number of papers analyzed in the third evidence collection.]{Overview of the evidence analyzed and filtered in each step of the evidence collection for comprehension measures.}
    \label{tab:evidence-search-results-3}
    \begin{tabular}{l r r r}
        \toprule
        \textbf{Source} & \textbf{Analyzed} & \textbf{Excluded} & \textbf{Final} \\
        \midrule
        a) Primary Papers & 95 & -88 & 7 \\
        b) Cited in Primary Papers & 3 & -3 & 0 \\
        c) Title Search & 53 & -51 & 2 \\
        d) Snowballing & 134 & -131 & 3 \\
        \midrule
        Evidence Profile & 12 & -5 & 7 \\
        \textbf{Total} & \textbf{285} & \textbf{-278} & \textbf{7} \\
        \bottomrule
    \end{tabular}
\end{table}

\begin{figure}[t]
    \centering
    \includegraphics[width=\linewidth]{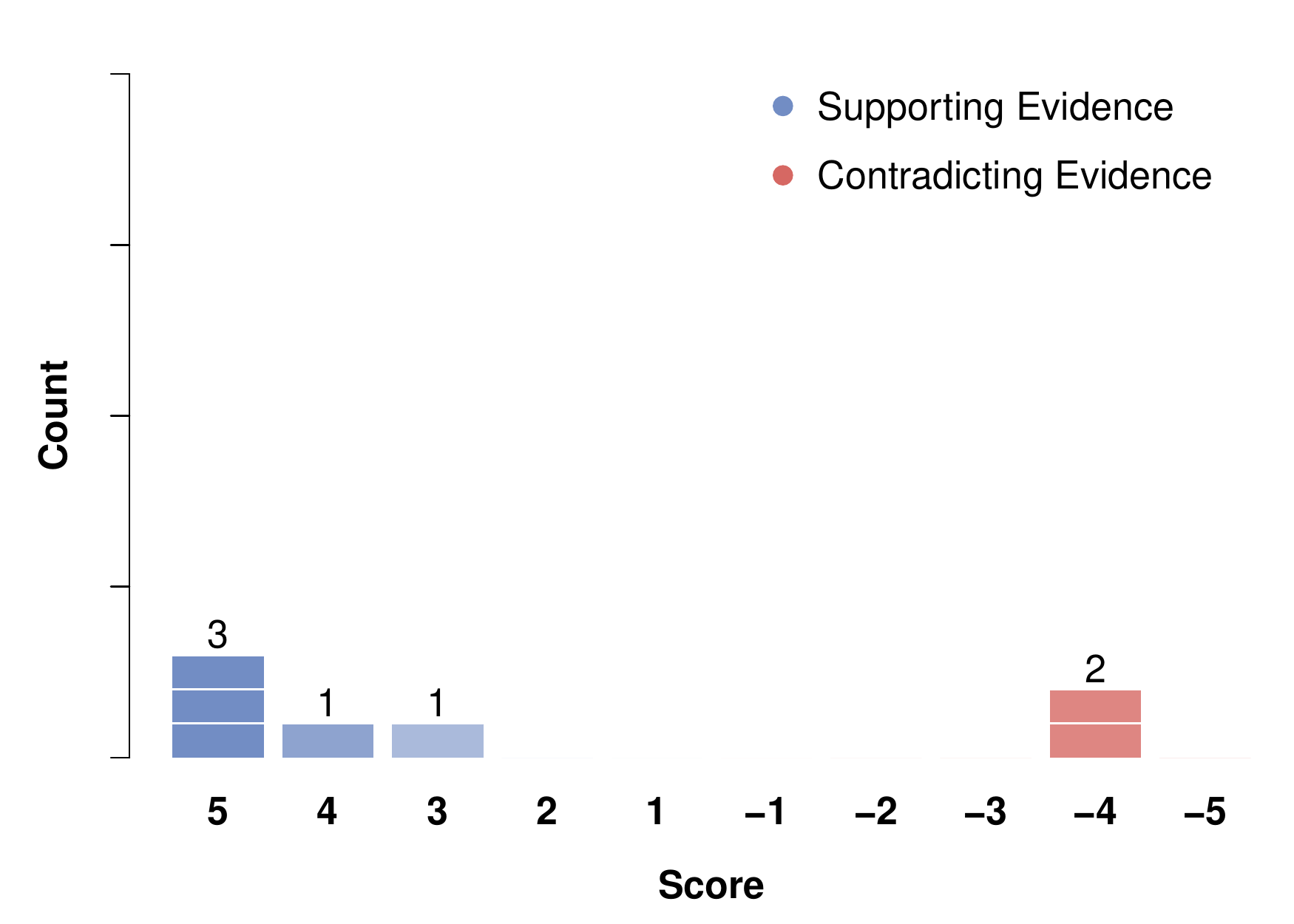}
    \caption{Evidence Profile 3: Different comprehension measures do not correlate with each other.}
    \label{fig:evidence-profile-3}
\end{figure}

After filtering, 12 papers remained and were evaluated as potential pieces of evidence using the evidence profile.
In this evaluation, 5 more papers were discarded as they did not meet the criteria to be considered evidence.
The final evidence profile is shown in \cref{fig:evidence-profile-3}.
Most evidence supported that the commonly used comprehension measures do measure distinct aspects of program comprehension and are not correlated.
But overall, only a small amount of evidence was found: 5 (71\%) pieces of evidence were rated as positive evidence and 2 (29\%) were rated as negative evidence.
We found 3 pieces of strong positive evidence, but no strong negative evidence.

\begin{tcolorbox}[colback=white,colframe=black, title={\textbf{\textcolor{colorE}{RQ2}: \textsf{Main Findings}}}]
    \begin{itemize}[leftmargin=*]
            \item \textbf{Programming experience}:\\ \textcolor{colorA}{37} positive and \textcolor{colorB}{12} negative evidence
            \item \textbf{Program length}:\\ \textcolor{colorA}{6} positive and \textcolor{colorB}{7} negative evidence
            \item \textbf{Comprehension measures}:\\ \textcolor{colorA}{5} positive and \textcolor{colorB}{2} negative evidence
            \item 94\% of cited evidence in the primary papers did not meet our criteria for evidence and was excluded.
            \item No evidence was categorized as strong negative evidence (-5).
    \end{itemize}
\end{tcolorbox}
\vspace{8pt}

\section{Discussion}
\label{sec:discussion}

We will first consider the results of the evidence profiles on the three most frequently discussed threats to validity.

Overall, the \textbf{first evidence profile} confirms that, often, programming experience does influence the comprehension performance of programmers.
In multiple cases, experienced programmers showed different comprehension behavior when compared to novices~\cite{Adelson.1984, Yu.2019} and this difference could be measured in their performance~\cite{Wiedenbeck.1985, Orlov.2017, Medeiros.2019b}.
In contrast, however, we also found credible evidence contradicting those assertions.
Twelve studies reached the conclusion that programming experience does not influence code comprehension.
One explanation for this contradiction could be the specific contextual factors of each study.
Siegmund et al.~\cite{Siegmund.2014} found that depending on how programming experience is measured and operationalized, its predictive power varies.
Moreover, some negative evidence still found correlations with very specific types of experience measures such as self-estimated Java knowledge~\cite{Peitek.2020} or correlations for only specific comprehension measures such as the number of eye fixations~\cite{Jessup.2021}.
In other cases, the range of programming experience was quite limited, for example due to only including students as participants~\cite{Peitek.2020, Hendrix.2002}.
These results imply that mentioning programming experience as potential threat alone is insufficient when publishing experiment results.
Researchers should instead discuss the threat within the context of their study and discuss how and why they adapted procedures, measures, and artifacts to mitigate it.

The \textbf{second evidence profile} indicates that an influence of program length on code comprehension behavior or performance should not be assumed blindly in every context.
We found conflicting evidence for this potential threat to validity.
Ribeiro et al.~\cite{Ribeiro.2018} discuss similarly conflicting evidence on program length in their study.
Even after conducting a follow-up study, they were unable to reach a clear conclusion.
Novices and experts had differing opinions regarding program length and its influence on comprehensibility and readability.
They also mention that differences in the procedures and tools used to measure lines of code may affect the comparability of results from different studies.
While the overall number of samples is small, we can identify some patterns when comparing positive and negative evidence.
All studies rated as positive evidence included students as participants, and one of them included both students and experts.
Every study that only included experts was rated negative.
This pattern suggests that while program length may influence the comprehension performance of novices, experts appear to perform consistently, regardless of program length.

The results of the \textbf{third evidence profile} mostly support the notion that the commonly reported comprehension measures are not correlated.
The three supporting pieces of evidence found no correlations between the time and accuracy of comprehension task performances, suggesting that they measure different effects, aspects, or dimensions of comprehension task difficulty~\cite{Iselin.1988, Binkley.2009, Ajami.2019}.
However, both pieces of negative evidence found correlations between the time and accuracy of comprehension tasks~\cite{Gilmore.1984, Hall.1986}.
Furthermore, the two studies that compared physiological measures with other measures found no correlation with subjective ratings~\cite{Yeh.2021} and time and accuracy of comprehension tasks~\cite{Fakhoury.2020, Yeh.2021}.
The significance of this finding becomes more apparent when one considers that of the 95 primary papers, more than one third (37) used only a single measure to assess the comprehension performance of participants.
Depending on the study context, this may cause them to miss aspects of code understanding that would have been uncovered if they had analyzed more than one measure.
If different proxy measures are associated with different aspects of program comprehension, this poses further difficulties for meta-studies of comprehension experiments.
For example, two studies might obtain different results because they measured comprehension performance in different ways and not due to other intrinsic factors.
Using the results from this work, we were unable to identify concrete patterns in the context factors which would suggest why comprehension measures correlate in some instances and not in others.
A dedicated follow-up study, solely focused on comparing different comprehension measures, may shed more light on this question.
For now, combining studies with different comprehension measures should be done with care until we better understand the way they measure different aspects of comprehension.

When compared, we found much less evidence overall for the influence of program length and the comparability of comprehension measures than for the influence of programming experience.
Even though program length and comprehension measures are the second and third most discussed threats to validity in code comprehension experiments, there are few studies examining their influence.
With the little evidence we found, we nevertheless arrived at similar results to the first evidence search:
The extent of all three validity threats is context-dependent.
The influence of a validity threat varies for each individual study and must, as such, be interpreted in the light of the contextual factors surrounding it.
When seeking patterns to explain the contradiction of evidence, we found that the way a threat impacts the result of a study can depend on how a confounding factor is measured, what the sample population is, and how the experiment is designed in general.

While this conclusion might sound obvious at first, we often find generic statements about a potential threat in primary studies.
For example, a study might solely mention that their experiment sample consisted of students without further elaboration.
Discussing a study characteristic this way, however, does not clarify why that design decision might constitute a threat to the validity of that specific study.
This issue is problematic regardless of whether references are used to back up assumptions, as generic threats are used in place of more nuanced discussions that take the study-specific context factors into account.
Evidence should be cited as an additional layer to further support the explanations.
For example, Jbara and Feitelson~\cite{Jbara.2017} show how evidence from past studies can be used when discussing the threats to validity of their study.
They describe how their use of undergraduate students was adequate because the students were at a high enough skill level to complete the experimental task and cite evidence to support their assertion.
In this way, they use evidence to guide the threat discussion within the context of their study.

\subsection*{Minor Observations}
Besides this main finding, we also made some further observations.
First, the studies used in the primary papers to support claims about threats to validity were, with one exception, almost entirely dismissed as evidence.
Studies were mostly excluded because they either did not relate to program comprehension, solely cited another mention of the threat, or they were not relevant at all.
This pattern was consistent over all three evidence searches.

Second, the results of our review of how threats to validity are reported match the investigation by Wyrich et al.~\cite{Wyrich:2022:40Years}.
In general, researchers in program comprehension tend to focus on threats relating to the code snippets and the study participants the most.
The factors we investigated in our evidence search, programming experience, program length, and comprehension measures are among the most frequently mentioned threats in both studies.
The terms used to describe these threats differ between the works.
For example, the category named \enquote{comprehension measures} can be found in the category \enquote{instrumentation} in Wyrich et al. and \enquote{program length} is named \enquote{program size}.
Looking at the bigger picture, the results for program comprehension studies are in line with reporting in SE in general. 
Siegmund et al. found that in SE, 51\% (47\% in our study) of papers discussed threats to validity in a dedicated section and 23\% (35\% in our study) differentiated between different types of validity~\cite{Siegmund.2015}.

Third, we found that, overall, there was less evidence for less common threats to validity.
This is in line with what would be expected intuitively, if fewer researchers deem that a threat poses a danger to a study's validity then corollary, fewer will investigate whether that assumption holds.
Meta-studies such as this one can highlight which threats are less frequently examined and provide directions for further experimentation.

Fourth, we observed that throughout the entire evidence collection process, not a single study was classified as a \enquote{\mbox{-5}}, which would represent strong negative evidence.
While we do not have a conclusive reason for this, it is common for researchers to experience difficulties publishing negative results.
Obtaining negative results can be disheartening, but publishing them is nonetheless crucial and provides value to the scientific community~\cite{Weintraub.2016, Paige.2017, Borji.2018}.

\subsection{On the Desire for Evidence}
Looking back at The Highway Shooter case, we also want to acknowledge some key differences between the criminal case and a scientific experiment.
First, in the media coverage following the case, the recording and usage of license plates as evidence was criticized as a breach of privacy.
In the context of academic research, similar issues with data privacy mainly arise when collecting data for primary research.
When synthesizing results of different independent studies, however, applying their research findings as evidence with proper citation is usually encouraged.
Furthermore, when creating evidence profiles, synthesis is even possible without disclosing the data sets of the studies.
The information contained in the paper itself is often sufficient to assess its weight as evidence.
Another important difference is that there was only one perpetrator responsible for the crimes, whereas in any given experiment, there may be more than one validity threat.

When designing an experiment, it is infeasible to conduct systematic reviews to collect evidence for each of the dozens of potential threats.
Rather, researchers may look into existing research summarizing evidence and then contextualize it by comparing the underlying studies with the characteristics of their own methodology.
The present paper can be one such resource for evidence collection, future meta-studies on common research questions another.
Evidence is then used both during the design stages of an experiment and when interpreting and discussing its results.

Conversely, one may discuss the nature and role of the threats to validity section in scientific literature itself.
While we examined and reported on various descriptive aspects of the threats reported in existing papers, we did not inquire into why researchers chose to report and discuss specific threats in a certain way.
Furthermore, our research does not provide explicit guidance on how a paper author ought to write a \textit{threats to validity} section.
While the position presented in this paper suggests that reported threats should be supported with evidence whenever possible, this is not necessarily a sentiment shared by all members of the scientific community.
A section on validity threats may also be a place where researchers should be able to speculate without concrete evidence and point out potential shortcomings as directions for future research.
This debate should be conducted within the respective research communities, and agreements should then be recorded and incorporated into existing guidelines for the reporting of validity threats.

\subsection{Implications}

The results presented in this work lead us to concrete suggestions for the larger research community.
First, threats to validity are dependent on individual context factors of a study.
Researchers should therefore explicitly discuss the influence of each threat within the context of their study.
Merely listing off potential threats alone is not sufficient.
Second, context discussions for validity threats should be supported by existing evidence, rather than relying on intuition or speculation.
Meta-studies provide the appropriate evidence base for this endeavor, as they analyze the influence of a threat in multiple different study contexts.

\subsection{Limitations}
Evidence profiles are created by qualitative evidence evaluation from human reviewers, and thus may contain bias.
To mitigate this threat, two researchers independently rated the evidence and then compared their results to reach an agreement.
The threat extraction, categorization, and subsequent evidence collection was done by only one researcher, which again might incur a bias and threaten the internal validity of our study.
However, when comparing the threat categories and number of occurrences with a previous systematic review on confounding factors in program comprehension studies by Siegmund and Schumann~\cite{Siegmund:2015:Confounding}, we find mostly similar results regarding threat codes and their frequencies.

Throughout this work, we used the terms program comprehension and code comprehension interchangeably for comprehension of source code, as is currently the norm.
However, in principle, we agree with Wyrich et al.'s observation that code comprehension studies form a subset of program comprehension studies~\cite{Wyrich:2022:40Years}.
The list of primary papers we built on was a result of a previous search by Wyrich et al. and its search parameters were not fully aligned to the goals of our work.
In their study, Wyrich et al. focused on a subset of program comprehension experiments, namely experiments on bottom-up code comprehension.
This means that one might obtain a different ranking for the most common threats in the broader area of program comprehension experiments, limiting the external validity of our results.
In the evidence collection, this limitation was mitigated as we added additional search sources.

When analyzing our data, we also found a rather strong bias in the number of studies with quantitative data from experiments over qualitative data.
Even though we did not explicitly exclude any papers in the evidence search on the grounds of them being a qualitative study, the results still heavily favored controlled experiments.
In numbers, 68 out of 69 (99\%) pieces of evidence in the evidence profiles were from experiments.
To avoid such bias in future studies, we recommend formulating descriptions of evidence profiles in a more method-neutral terminology (e.g., avoiding terms such as confounding factors, which are primarily indicative of experiments).
On a related note, new evidence profiles for validity threats from code comprehension studies using other research methodologies than experimentation can, provided equal or similar epistemological stances, help similarly well in making informed design decisions.
It depends on one's epistemological standpoint and understanding of the concept of evidence to determine whether it is appropriate to use evidence in the design of non-experimental studies.

When developing the evidence profiles, we decided to focus on threat categories instead of individual codes.
For example, rather than searching for evidence of programming experience as threat to internal or external validity, we combined the two into the category of programming experience.
A finer distinction was not necessary for our purpose, as evidence for the influence of programming experience on code comprehension is equally relevant for all validity types.
Further, a finer distinction would have been difficult to make.
As previously noted, validity threats in primary studies are currently not discussed in a way that makes clear what type of validity the authors believe their study characteristics would affect.

Another limitation in the external validity of our study is that due to the amount of effort required to systematically collect evidence, we were only able to analyze three of dozens of different validity threats.
We focused on the most frequently mentioned threats, as we expected them to provide the most value for the widest range of program comprehension studies.
It is possible that repeating the same methods for the remaining threats will uncover new idiosyncrasies of program comprehension.
As such, generalizing the results obtained in this work to other threats should be done with care.

\section{Conclusion}
\label{sec:conclusion}

In this work, we investigated threats to validity in program comprehension experiments.
First, we analyzed the state of the art by reviewing the reported validity threats in 95 papers of primary research.
We found that while most studies mentioned threats to validity, few supported them with corresponding evidence.
Furthermore, only in one case did the cited evidence support the validity of a threat and meet our quality criteria to be included in an evidence profile.
Next, we searched for evidence regarding the three most frequently mentioned threat categories.
Our evidence collection yielded both positive and negative evidence for each threat category, seemingly leading to divergent results.

After looking closer into the collected evidence and comparing the characteristics of the different studies, we concluded that validity threats are highly context-dependent.
Even the threats that are intuitively expected to affect program comprehension, such as programming experience, depend on how they are measured, the sample population, comprehension tasks, and other context factors.
Therefore, we must consider all individual characteristics of a study when assessing potential validity threats to develop methodologies that use evidence as a basis for implementing appropriate mitigation techniques.
Furthermore, discussions of validity threats in papers need to explicitly address context factors and researchers should use evidence to support these discussions.
Finally, we encourage the usage of threat catalogs and recommend the adherence to reporting guidelines for threats to validity to improve the reproducibility and comparability of study results.

Structured guidelines for reporting threats to validity need to be established further to inform researchers on how they can incorporate evidence into their validity assessments.
We need more knowledge documentation on which threats exist, the evidence supporting them, the context in which they occur, and which mitigation techniques can be used to address them.
Previous works laid the groundwork in this endeavor by documenting threats in software engineering studies in a knowledge base and providing guidelines for controlling the influence of confounding factors in program comprehension experiments, respectively~\cite{Biffl.2014, Siegmund:2015:Confounding}.
These works can be further extended by incorporating evidence and documenting threats to validity in a common database.

We envision a future in which scientists regularly apply our methodology to collect and summarize the evidence for additional threats to validity and for experiments far beyond code comprehension.
Even when looking at just three of the most common threats, we found less evidence for less popular threats, which might indicate even bigger gaps in evidence for those threats that have yet to be analyzed.
Moreover, our approach may be applied to other domains to investigate how threats to validity affect experiments there.

Taking a final look back at the past, in 2014, The Highway Shooter was sentenced to 10 years in prison.
Good investigative work led to the necessary evidence and ultimately to the apprehension of the criminal.
Evidence profiles in software engineering research equally lead to necessary data points, not to catch criminals, but, in our case, to gain more certainty about the actual impact of threats to validity in program comprehension experiments.

\section{Data Availability}
\label{sec:data_availability}
We publish our dataset on Zenodo for transparency and reproducibility of our approach~\cite{replication:package}.
The dataset comprises data that emerged in intermediate steps of the validity threat analysis. This includes extracted text passages from the primary studies and complete lists of threat codes, threat categories and threat themes. Further, the dataset contains artifacts produced during the evidence search and the creation of the evidence profiles. We document comprehensibly, for example, how individual reviewers evaluated individual pieces of evidence and reached a final agreement. Finally, R scripts are included, that mainly deal with quantitative analyses and plotting.

\section*{Acknowledgments}
We thank members of the German Federal Criminal Police Office (BKA), who wish to remain anonymous, for their information and guidance on The Highway Shooter case, which inspired us when presenting our findings.
We are very grateful to three anonymous reviewers, whose feedback helped to improve the paper.
We also thank Katharina Plett for proofreading to enhance the text quality.

\bibliographystyle{IEEEtran}
\bibliography{main}

\end{document}